\documentclass[sigconf]{acmart}
\AtBeginDocument{%
  }

\copyrightyear{2026}
\acmYear{2026}
\setcopyright{cc}
\setcctype{by}
\acmConference[ICSE-SEIP '26]{2026 IEEE/ACM 48th International Conference on Software Engineering}{April 12--18, 2026}{Rio de Janeiro, Brazil}
\acmBooktitle{2026 IEEE/ACM 48th International Conference on Software Engineering (ICSE-SEIP '26), April 12--18, 2026, Rio de Janeiro, Brazil}
\acmDOI{10.1145/3786583.3786890}
\acmISBN{979-8-4007-2426-8/2026/04}

\usepackage{xcolor}
\usepackage[dvipsnames]{xcolor}
\usepackage{todonotes}
\usepackage{graphicx}
\usepackage[table,xcdraw]{xcolor}
\usepackage{enumitem}
\newcommand{\review}[2]{\textcolor{black}{#2}}
\begin{document}

\title["Where is My Troubleshooting Procedure?"]{"Where is My Troubleshooting Procedure?": Studying the Potential of RAG in Assisting Failure Resolution of Large Cyber-Physical System}


\author{Maria Teresa Rossi}
\affiliation{%
  \institution{University of Milano-Bicocca, Italy}
  \country{Gran Sasso Science Institute, Italy}}
\email{maria.rossi@unimib.it}

\author{Leonardo Mariani}
\affiliation{%
  \institution{University of Milano-Bicocca}
  \city{Milan}
  \country{Italy}}
\email{leonardo.mariani@unimib.it}

\author{Oliviero Riganelli}
\affiliation{%
  \institution{University of Milano-Bicocca}
  \city{Milan}
  \country{Italy}}
\email{oliviero.riganelli@unimib.it}

\author{Giuseppe Filomeno}
\affiliation{%
  \institution{University of Milano-Bicocca}
  \city{Milan}
  \country{Italy}}
\email{g.filomneo@campus.unimib.it}

\author{Danilo Giannone}
\affiliation{%
  \institution{University of Milano-Bicocca}
  \city{Milan}
  \country{Italy}}
\email{d.giannone2@campus.unimib.it}

\author{Paolo Gavazzo}
\affiliation{%
  \institution{University of Milano-Bicocca}
  \city{Milan}
  \country{Italy}}
\email{p.gavazzo@campus.unimib.it}

\renewcommand{\shortauthors}{Rossi et al.}

\begin{abstract}
In today's complex industrial environments, operators must often navigate through extensive technical manuals to identify troubleshooting procedures that may help react to some observed failure symptoms. These manuals, written in natural language, describe many steps in detail. Unfortunately, the number, magnitude, and articulation of these descriptions can significantly slow down and complicate the retrieval of the correct procedure during critical incidents. Interestingly, Retrieval Augmented Generation (RAG) enables the development of tools based on conversational interfaces that can assist operators in their retrieval tasks, improving their capability to respond to incidents. 

This paper presents the results of a set of experiments that derive from the analysis of the troubleshooting procedures available in Fincantieri, a large international company developing complex naval cyber-physical systems. Results show that RAG can assist operators in reacting promptly to failure symptoms, although specific measures have to be taken into consideration to cross-validate recommendations before actuating them. 
\end{abstract}





\maketitle

\section{Introduction}\label{sec:intro}

Large industrial Cyber-Physical Systems (CPS) consist of many heterogeneous components (e.g., software, hardware, and mechanical components) that have to interact properly to guarantee the intended functionality \cite{baheti2011cyber}.
However, their inherent complexity often leads to unexpected behaviors or failures, and operators have to intervene promptly to solve or work around issues~\cite{campos2025predicting}.

Operating in a timely and proper manner in response to misbehaviors can be extremely challenging. In fact, operators may have to carefully pair failure symptoms with many different procedures that can potentially be actuated. For instance, the industrial system we considered in this paper is operated by Fincantieri, our partner in the ATOS project ({\small \url{https://sites.google.com/unimib.it/atos/home}}), and involves over 100 manuals with hundreds of procedures described in each of them. Depending on the symptoms, there might be hundreds of troubleshooting procedures that operators may potentially activate to address an observed problem.

This problem is exacerbated by the fact that many failure symptoms are common across multiple troubleshooting procedures. For example, in the troubleshooting procedures we experimented with, a loose or uneven pipe connection may occur in both the installation and the resolution of pressure issues of the fire prevention system. Similarly, drips or leaks are present in procedures related to both valve malfunction and loose connections.

Additionally, these procedures are documented in natural language, lacking a formal machine-processable specification.
For instance, the troubleshooting procedures used in our experiments are documented as XML files, where the XML elements are used to distinguish the trigger of a troubleshooting procedure (i.e., the set of failure symptoms that justify the execution of the procedure) and the description of the procedure itself. However, all descriptions are in plain natural language. Effectively browsing and searching within this corpus of information can be challenging, and even the most experienced operators may react slowly or incorrectly due to the difficulty of this task. 

These challenges are not unique to Fincantieri. Troubleshooting manuals of large CPS are often documented with natural language documents that are hard to parse and process~\cite{mertz2024challenges, Cunningham2022,lovis2015troubleshooting}.

An intuitive and useful option for operators is using natural language interfaces to request the appropriate procedure with a natural language description of the observed failure symptoms. For instance, operators may want to simply ask questions such as "\textit{What should I do if the pressure of subsystem X is higher than Y and the temperature alarm has been raised for subsystem Y?}". 

\begin{table*}[th]\begin{small}
\resizebox{0.75\textwidth}{!}{%
\begin{tabular}{|l|l|l|l|}
\hline
\textbf{ID} & \textbf{Failure Symptom} & \textbf{Possible Cause} & \textbf{Troubleshooting Action} \\ \hline
11-A & \begin{tabular}[c]{@{}l@{}}During operation of E/pump and its \\ desalinator on the automation consoles, \\ the vibration values for one or more \\ accelerometers are not present\end{tabular} & \begin{tabular}[c]{@{}l@{}}The signal cable from the accelerometer \\ panel to the ship's automation system is \\ not properly connected or is damaged.\end{tabular} & \begin{tabular}[c]{@{}l@{}}Check that the connection cable from \\ the accelerometer panel to the automation \\ system is correctly connected and not \\ damaged. Restore if necessary\end{tabular} \\ \hline
11-B & \begin{tabular}[c]{@{}l@{}}During operation of E/pump and its \\ desalinator on the automation consoles, \\ the vibration values for one or more \\ accelerometers are not present\end{tabular} & Accelerometer is faulty & Replace the accelerometer \\
\hline
8-A & \begin{tabular}[c]{@{}l@{}}Pressing the E/P start button on the \\ desalinator control display does not \\ start the E/P of the E/Pump Module.\end{tabular} & \begin{tabular}[c]{@{}l@{}}Desalinator panel touch screen\\ failure.\end{tabular} & Replace faulty touch screen \\ 
\hline
\end{tabular}%
}
\caption{Example of the key items present in the troubleshooting manuals.}
\label{tab:troub_example}\end{small}
\end{table*}

So far, research has mostly focused on the design of repair techniques~\cite{8089448,10.1145/3631974} and automatic workarounds~\cite{10.1145/2755970}, with limited attention on the design of semi-automatic solutions that can yet improve the capability to respond to system failures. 
With the development of Large Language Models (LLMs), natural language interfaces are gaining popularity~\cite{Zhang2024}.
In particular, the \textit{Retrieval-Augmented Generation (RAG)} architecture~\cite{martineau2023retrieval} can promisingly leverage \emph{semantic distances} and \textit{LLM}s  to \textit{search \& retrieve} information from a \emph{knowledge base} populated with documents (e.g., the documents with troubleshooting procedures) that must be used to answer the asked questions (e.g., questions about how to react to failures). 
In \cite{Rossi2025}, we report preliminary results of a RAG-based system that automatically generates prompts
from anomaly data and retrieves relevant troubleshooting procedures from system documentation.

This appealing option is not, however, free of challenges and perils. Indeed, responses are useful only as long as they are accurate.
To understand the feasibility and effectiveness of this solution, \emph{we empirically studied} to what extent RAG can be used by operators to efficiently \emph{retrieve} troubleshooting procedures from natural language queries. 
To achieve this goal, we investigated five research questions that overall required the assessment of queries.

\noindent \textbf{RQ1-Accuracy}: \emph{To what extent can a RAG recommend accurate troubleshooting instructions?} With this research question, we study both the capability of retrieving the correct troubleshooting procedures from the knowledge base and the capability of formulating an adequate natural language description of the operations that the operator must complete to respond to an incident.

\noindent \textbf{RQ2-Sensitivity}: \emph{To what extent is the accuracy of the RAG dependent on how the questions are formulated?} RQ2 studies how the level of agreement between the content of the question and the content of the knowledge base impacts the accuracy of the results. In fact, questions might be formulated imprecisely, or not using the same terms present in the knowledge base. This RQ assesses its impact, both in the retrieval and in the formulation of the response. 

\noindent \textbf{RQ3-Derivation}: \emph{To what extent is the RAG able to derive undocumented troubleshooting procedures?} This RQ investigates whether the generative capabilities of the LLMs can be useful to guess procedures that are not explicitly documented in the KB, supporting operators towards the resolution of undocumented problems. 

\noindent \textbf{RQ4-Qualitative Assessment}
\emph{To what extent do the results produced by the best-performing configurations reflect accurate and complete troubleshooting procedures?}
This RQ complements RQ1, RQ2, and RQ3, which, because of their scale, rely on automatic and potentially imprecise evaluation methods, with a precise assessment based on the manual inspection of the responses generated by the best-performing configuration. The goal is to report and discuss the results qualitatively, identifying whether the returned procedures include all the troubleshooting steps or introduce inaccuracies.

\textbf{RQ5-Performance}: \emph{How quickly can the RAG respond to the questions?} RQ5 studies how quickly RAG can respond to the users' questions, to investigate their suitability as tools available to operators to handle incidents, as soon as they occur.

Our results reveal that a RAG can \emph{quickly suggest useful procedures}, even \emph{accommodating for some imprecision} on the way questions are formulated by operators. On the other hand, the sources used by a RAG to generate a response \emph{must always be returned} together with the answer, so that operators can cross-validate recommendations, if necessary. Sometimes, the RAG has also been able to \emph{derive undocumented procedures} from the information stored in the knowledge base, helping operators with unexpected scenarios.
From our experiments, a \emph{trade-off between speed and accuracy} of answers in relation to model size emerges. 
Indeed, smaller language models offer significant advantages in terms of computational costs, while larger models foster accuracy of the answers by decreasing response times. 
The encouraging results pave the way to more research in the area, to understand how emergencies can be addressed more efficiently with the help of LLMs, especially in the frequent situation where the procedures are encoded in natural language. Our lesson learned identifies several possible areas of improvement for future research on this subject.
\section{Natural Language Troubleshooting Procedures}\label{sec:problem}

Technical manuals are essential for diagnosing and resolving system failures. However, quickly identifying the correct troubleshooting procedure within a large corpus of documents can be difficult and time-consuming. Even the most experienced operators are not used to failures, which occur rarely, and thus have to rely on the troubleshooting procedures documented in manuals to react to them.

Currently, operators manually search through these documents, primarily using the \textit{search} feature available in the editors. They must first locate the appropriate document and then browse or search within it to find a suitable troubleshooting procedure. This process is inefficient and requires familiarity with the manuals, increasing the risk of errors, such as missing the correct procedure or selecting the wrong one.  

The manuals available in Fincantieri are similar to the manuals that can be found in many other contexts: they contain natural language descriptions of symptoms and related troubleshooting procedures. Table~\ref{tab:troub_example} shows an example of the key items reported in the manuals. Specifically, for each \emph{failure symptom}, a list of \emph{possible causes} and their corresponding \emph{troubleshooting actions} are provided. For instance, we reported procedures \emph{11-A} and \emph{11-B} that are related to the same failure symptoms and that propose two different troubleshooting procedures depending on the possible cause of the failure. Instead, the procedure \emph{8-A} comes from a different document, but it is related to a failure symptom dealing with a component called \textit{E/pump} as for the other two procedures. This exemplifies how a trivial search for symptom names would lead to the selection of multiple alternative procedures that could be activated, with little clue for the actual choice.

\begin{figure}[htb!]
    \centering
    \includegraphics[width=\linewidth]{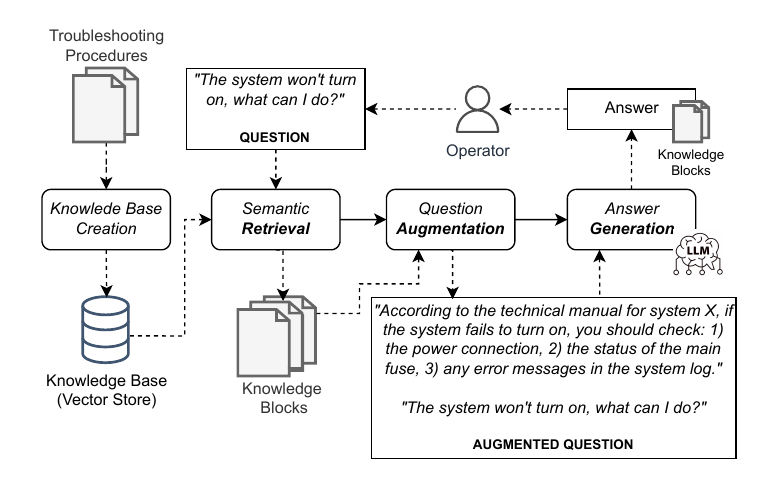}
    \caption{Architecture of the proposed RAG for the recommendation of troubleshooting procedures. 
    }
    \label{fig:architecture}
\end{figure}

\section{Troubleshooting with RAG} \label{sec:RAG}

A RAG allows users to ask LLMs natural language questions that need information stored in a Knowledge Base (KB) to be answered. 
The answers of the RAG shall consist of procedures that are then suggested to the operators. 
Note that operators are not supposed to apply procedures blindly. On the contrary, they use their skills to decide about their suitability. The RAG supports cross-validation of the responses, since each answer is associated with the items of the KB that motivate the response, offering the operators means to verify the answer by accessing the sources. 

Figure~\ref{fig:architecture} illustrates how we use the RAG to retrieve troubleshooting procedures of a naval CPS. 
The process starts with the step \emph{knowledge base creation} that parses the \emph{Troubleshooting Procedures} and embeds them into a \emph{Knowledge Base}, accessible to the \emph{LLM}.

The knowledge base is populated by first dividing the input documents into chunks (i.e., non-overlapping sections of text of fixed or variable length tagged with indices to facilitate information retrieval). These chunks serve as the basic units of knowledge accessible to the LLM and are then uploaded into the knowledge base~\cite{gao2023retrieval}.

Each chunk is then mapped into an embedding to allow for the semantic retrieval of the information; that is, the embedding maps chunks that are semantically related in nearby regions of the space, enabling effective similarity-based retrieval.  
Once the chunks have been uploaded, the knowledge base is ready to be used.

The user interested in retrieving the procedure to be applied in reaction to an issue asks a question that is submitted to the RAG. Before the question is routed to the LLM, the \emph{knowledge retrieval} looks for chunks that are semantically close to the question according to the cosine similarity~\cite{xia2015learning}. The intuition is that text (i.e., chunks) semantically close to the text in the question is likely useful to answer the question (e.g., it includes information about the problem described in the question and the possible reaction to that problem). 

In principle, semantic retrieval might already be sufficient to respond to the operators' questions: that is, operators may simply process the list of retrieved chunks one after the other. 
However, our experiments show that retrieval is difficult and often imprecise, due to the similarity between troubleshooting procedures, with the correct text often not reported in the first position, or not retrieved at all. The LLM used to process the output of the retrieval can, interestingly, compensate for this weakness.
The retrieved chunks are automatically embedded as a preamble in the prompt of the LLM.  
This helps the language model to ground its answer on the retrieved information, while exploiting generative capabilities to compensate for any inaccurate retrieval.

For instance, a user's question such as \emph{The system won't turn on, what can I do?} can be augmented with the retrieved chunks, obtaining a final augmented question that includes in the context a description such as \emph{According to the technical manual for system X, if the system fails to turn on, you should check: 1) the power connection, 2) the status of the main fuse, 3) any error messages in the system log file}. This step is particularly important to let the LLM have access to information about the system that is malfunctioning and the specific procedures that can be actuated. In fact, this information, stored in private documents of companies, would not be otherwise available to the model.

In the \emph{answer to the question} step, the augmented question is passed to the LLM that formulates the final response.

We designed the RAG to work with a fully local deployment, not relying on any third-party service, to preserve the intellectual property and the privacy requirements required by companies, thus not sending any information to any online service. For this reason, we have not considered the usage of online LLMs, such as ChatGPT~\cite{chen2024benchmarking} or Gemini~\cite{zhu2024accelerating}, and the project material is not provided in this paper.

To simplify the integration of multiple LLM models within the RAG, we used Ollama ({\small{\url{https://ollama.com/}}}) as the container of our RAG architecture. 
We used LangChain ({\small \url{https://www.langchain.com/}}) as RAG framework and Chroma ({\small \url{https://www.trychroma.com/}}) as vector store, since they are both popular frameworks offering production-quality capabilities.

\section{Methodology} \label{sec:methodology}

To evaluate the effectiveness of RAG in supporting troubleshooting activities in large industrial Cyber-Physical Systems, we designed a set of experiments guided by five main research questions. Each question targets a specific aspect of the troubleshooting identification process:

\noindent \textbf{RQ1 – Accuracy}: \emph{To what extent can a RAG produce accurate troubleshooting instructions?}  
    This question investigates both the precision of the retrieval task and the quality of the generated response. It is further divided into:
    
        \indent \textbf{RQ1.1 – Retrieval Accuracy}: \emph{Can a RAG retrieve troubleshooting procedures from operators' queries?} This sub-RQ specifically assesses the precision of the retrieval task of the RAG, namely the output of the \emph{Semantic Retrieval} task in Figure~\ref{fig:architecture}, by measuring if and in which position the block with the correct troubleshooting procedure occurs in the set of retrieved blocks.
        
        \indent \textbf{RQ1.2 – Response Accuracy}: \emph{Can the LLM exploit the retrieved information to accurately describe the troubleshooting procedure?} This sub-RQ studies if the LLM can properly formulate the troubleshooting procedure that must be applied to work around a problem.  In fact, the role of the LLM might be either beneficial, producing an accurate description of the troubleshooting procedures from the blocks returned by the retrieval task, or deleterious, producing an inaccurate response from the retrieved blocks. Ultimately, the accuracy of the response (the answer block in Figure~\ref{fig:architecture}) determines the accuracy of the whole RAG system. To scale the evaluation to a large set of responses obtained with different questions, models, configuration, and repetitions, we adopt an LLMs-as-jury strategy, presented in Section~\ref{sec:judges}.
    
    \noindent \textbf{RQ2 – Sensitivity}: \emph{To what extent is the accuracy of the RAG dependent on how the questions are formulated?}  This question studies how the level of agreement between the content of the question and the content of the knowledge base impacts the accuracy of the results. In fact, questions might be formulated imprecisely, or not using the same terms present in the knowledge base. This RQ assesses its impact, both in the retrieval and in the formulation of the response. It is split into:
    
        \indent \textbf{RQ2.1 – Retrieval Sensitivity}: \emph{How sensitive is the retrieval to the way questions are formulated?} This sub-RQ studies the sensitivity of the retrieval task to the input questions. We assess the results using the same metrics used for RQ1.1.
        
        \indent \textbf{RQ2.2 – Response Sensitivity}: \emph{How sensitive is the LLM to the way questions are formulated?} This sub-RQ studies the sensitivity of the final response to the input question. We assess the results using the same metrics used for RQ1.2.
    
    \noindent \textbf{RQ3 – Derivation}: \emph{To what extent is the RAG able to derive undocumented troubleshooting procedures?} LLMs have generative capabilities, that is, they can suggest procedures that are not necessarily in the knowledge base, but they could potentially derive new procedures, exploiting both the knowledge present in the knowledge base and the knowledge intrinsically stored in the LLMs themselves. This sub-RQ assesses the capability of the RAG to recommend undocumented troubleshooting procedures that have not been anticipated by the domain experts, and thus are not present in the manuals. To measure how diverse the answers are when the support of the KB is lacking, we compute the BLEU~\cite{haque2022semantic} metric between the expected and given response. 

\noindent \textbf{RQ4-Qualitative Assessment}
\emph{To what extent do the results produced by the best-performing configurations reflect accurate and complete troubleshooting procedures?}
To scale the evaluation to a large number of queries, RQ1-RQ3 use automatic metrics. However, these metrics only provide an approximation of the real accuracy of the answers. 
To gain a more accurate view of the results, we answer RQ4 by manually inspecting the outputs generated by the best-performing configurations in RQ1-3. The goal is to report and discuss the results qualitatively, identifying whether the returned procedures include all the troubleshooting steps or introduce inaccuracies.

    \noindent \textbf{RQ5 - Performance}: \emph{How quickly can the RAG respond to questions?}  This research question investigates the efficiency of the RAG, to assess the possibility of establishing actual live conversations between the operator and the RAG system.

\subsection{Questions Asked}
\label{sec:domandetestavanzate}

To sample a range of different cases, we randomly selected 25 distinct sets of symptoms and corresponding troubleshooting procedures from the available documents. 
For each selected set of symptoms, we defined four questions, all expecting the same troubleshooting procedure as an answer. Each question is characterized by a different level of precision in the included statements and the possible inclusion of contextual information. In total, we thus considered 100 questions. 

In particular, each question consists of two parts. The initial part is the question itself. It is a sentence that begins with "\textit{What should I do if...}" and continues with a description of the observed symptoms. The description could be either using the same terminology used in the troubleshooting (accurate question) or using an imprecise terminology (inaccurate question). To obtain the inaccurate question, we substitute the technical terms used in the documentation with more generic synonyms. The second part of the question is a reference to the document where the searched procedure is present. The second part is included for questions with context, and skipped for questions without contexts. We thus ask questions like \textit{"What should I do if the red indicator 'LOW PUMP PRESSURE' is on?"}.

The four types of questions capture four different use cases: the \emph{accurate questions with contexts} correspond to an operator with good knowledge of the available documentation who knows which document describes the searched troubleshooting procedure; the \emph{accurate questions without context} correspond to an operator who knows the domain terminology, but does not have the exact knowledge of the many documents available; the \emph{inaccurate questions with context} correspond to an operator who does not know the domain, but knows the documentation (e.g., a beginner who is not yet confident in the domain but has worked on the documentation long enough to know where to search); and the \emph{inaccurate question without context} corresponds to an operator who is not able to use the exact domain terminology used in the documentation and cannot provide a priori reference to the documentation that includes the answer to the question.

\smallskip

To specifically answer RQ3, we need to assess questions that have no direct answer in the KB. To this end, we initialized the knowledge base with three different configurations, which correspond to three cases with gradually less information stored. Given a question Q about some failure symptoms $S$ that must be handled with the troubleshooting procedure $T$, the \emph{no response} configuration corresponds to the case in which $S$ occurs in the KB, but not $T$; the \emph{no entry} configuration corresponds to the case in which both $S$ and $T$ do not occur in the KB; and the \emph{no KB} configuration corresponds to the case the KB is empty, and thus even the symptoms and troubleshooting procedures similar to $S$ and $T$ are missing. We generate all these configurations for the questions studied in the paper.

\subsection{Setup} \label{sec:setup}

\emph{Environment}. For the experiments, we run the application on a virtual machine via Azure (a \textit{\href{https://learn.microsoft.com/en-us/azure/virtual-machines/sizes/gpu-accelerated/ncast4v3-series?tabs=sizebasic}{Standard NC4as T4 v3}}) that use an Nvidia Tesla T4 GPU, with 16GB of VRAM, and an AMD EPYC 7V12 CPU.

\begin{table}[ht]
\resizebox{\columnwidth}{!}{%
\begin{tabular}{|l|l|l|l|l|}
\hline
\textbf{Model} & \textbf{Parameters} & \textbf{Correct} & \textbf{Correct Exceeding} & \textbf{Incorrect or}   \\ 
 & &  & \textbf{Expectations} & \textbf{Incomplete}  \\ \hline
\textit{Llama}& 3B & 2 & 0 & 2  \\ \hline
\textit{Mistral}& 7B & 1 & 1 & 2  \\ \hline
\textit{Qwen}& 7B & 1 & 2 & 1  \\ \hline
\textit{Phi}& 3B & 0 & 0  & 4  \\ \hline
\textit{Orca-mini}& 3B & 0 & 0 & 4  \\ \hline
\textit{Zephyr}& 7B & 0 & 0 & 4  \\ \hline
\end{tabular}%
}
\caption{Results of the early assessment of the LLMs.}
\label{tab:initial_test}
\end{table}

\emph{LLMs}. To identify the LLMs to be used for the experiment, we performed an early assessment of six LLMs of different sizes and nature, as summarized in columns \emph{model} and \emph{parameters} of Table~\ref{tab:initial_test}.
We formulated four questions, considering four different topics and phrasing the questions differently, to explore how variations in structure, tone, and contextual detail may influence the clarity and specificity of the required troubleshooting procedure. 
Since we formulated the questions based on the available troubleshooting document, we also know the answer to each question.

To assess the answers provided by the LLMs to the questions, each answer has been independently inspected by two authors, distinguishing \emph{correct answers}, which provide instructions equivalent to the troubleshooting procedures that must be applied to the considered case; \emph{correct answers that exceed expectations}, which add relevant and accurate information extracted from the KB to the correct answer; and \emph{incorrect or incomplete answers}, which do not fully describe the troubleshooting procedures that must be applied to the considered case.
If the two inspectors disagree on the assessment of an answer, they analyze the answer and discuss until they find an agreement.

Table~\ref{tab:initial_test} summarizes the results for this initial assessment. We can state that Llama\footnote{\url{https://www.llama.com/}}, Mistral\footnote{\url{https://mistral.ai/}}, and Qwen\footnote{\url{https://chat.qwen.ai/}} clearly provided the highest number of correct answers and the least number of incorrect answers. While \footnote{\url{https://learn.michttps://huggingface.co/HuggingFaceH4/zephyr-7b-betarosoft.com/it-it/windows/ai/models/get-started-models-genai}}, Orca-mini\footnote{\url{https://huggingface.co/pankajmathur/orca_mini_3b}} and Zephyr\footnote{\url{https://huggingface.co/HuggingFaceH4/zephyr-7b-beta}} struggled with the considered task, only providing inaccurate answers.

Based on these results, we select for the larger experimental campaign described in the next section Qwen (the best performing model), Mistral (the second best performing model), and Llama (the best non 7B parameters model).

\emph{LLMs' Configuration}. Regarding the temperature parameter, which affects the creativity of the model, we used the default value.
Regarding top-p (i.e., nucleus sampling~\cite{Holtzman:TopP:ICLR:2020}), which is a decoding parameter in the range $[0,1]$ that controls the randomness of the language model’s output by limiting the token sampling to the smallest possible set of tokens whose cumulative probability exceeds the specified threshold, we considered three values: 0.2, 0.5, 0.9. The choice of the values derives from the goal of sampling the range of possible values, without hitting extreme cases (e.g., top-p equals 0 or 1).

For the chunk size (i.e., the size of the retrieved blocks), we considered three values: 400, 800, and 1000. The chunk sizes refer to the number of characters into which the documents are split before being embedded and indexed. We chose these values to study how different granularity levels in the retrieval process affect the quality of the generated answers.

The retrieval returns four blocks that are used to augment the question. Four blocks represent a good tradeoff between excessive information retrieved and the probability of retrieving the correct block, as confirmed in our experiments.

\begin{figure}[htb]
    \centering
    \includegraphics[width=\linewidth]{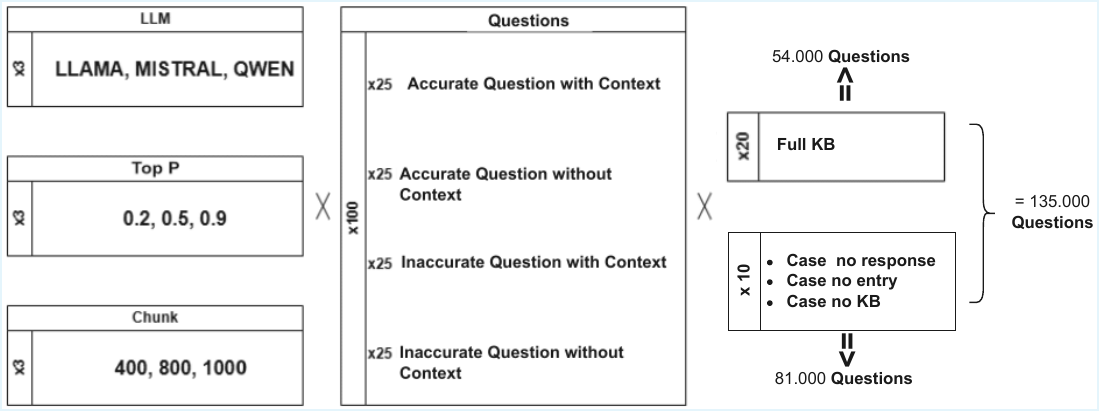}
    \caption{Total number of questions submitted to the RAG in our evaluation.}
    \label{fig:questions}
\end{figure}

\emph{Experiment space}. We study how all the parameters interact, as summarized in Figure~\ref{fig:questions},  
for a total of 135,000 questions assessed in this study: three LLMs, times three values of top-p, times three chunk sizes, times 100 questions, times 25 questions per category, and 20 repetitions with the full KB and 10 for the three cases with partial or no KB. 

We use the answers to the Accurate Questions with context to answer RQ1. We use the answers to the alternative formulation of the questions to answer RQ2. We use the questions without a direct answer in the KB to answer RQ3. 

\subsection{Evaluation With the Judges} \label{sec:judges}
Large-scale experimentation requires automated assessment of the answers. For each question we ask, we know by design what the correct answer extracted from the available documentation is. However, the same answer can be stated in many different but equivalent ways. We thus adopted an LLM-as-a-jury paradigm~\cite{hu2025training} and selected three LLMs, distinct from the ones used in the RAG to avoid any bias, as judges for the answers. The three LLMs are Gemma ({\small{\url{https://ai.google.dev/gemma?hl=it}}}), Granite ({\small{\url{https://www.ibm.com/granite}}}) and Hermes ({\small{\url{https://nousresearch.com/hermes3/}}}).
We selected these models since they perform well in evaluation and classification tasks, and their diversity in model architecture and training data ensures a more robust and unbiased judgment.

Each judge LLM is queried using the following template:
{\sc "The text within the brackets is referred to as the 'reference' (" + [reference] + " ). The text after the exclamation mark symbol is referred to as the 'comparison'. Your task is to assess the similarity between the content of the 'reference' and the content of the 'comparison', answering only with a number between '1' and '10' inclusive. If you answer '1', the content of 'reference' is completely different from the content of 'comparison', whereas if you answer is '10' the content of the 'reference' and the content of the 'comparison' are extremely similar. The check you have to make is not about the form, but about the content. It is not necessary that the exact same words are used, but the meaning must be the same. !" + [LLM response]}.

To decide how to employ the LLM judges, we first performed a small-scale quality control experiment to assess the judges in isolation and as an ensemble. In particular, we select 33 questions of various complexity and length, and submit the questions and the responses formulated by the RAG to the three LLM judges. 

To produce the ground truth, two of the authors manually assessed the same set of responses with a score within the same scale used by the judges and then compute the correlation between the judges' scores and the ground truth scores to determine the best combination. We considered the three judges in isolation, and the average and mean scores of the responses produced by the three judges as a possible option. Table~\ref{tab:judges} summarizes the results.

\begin{table}
\begin{center}
\begin{tabular}{l | c | c | c | c | c}
\hline
\textbf{Judge} & Gemma & Granite & Hermes & Avg & Mean  \\
\hline

\textbf{Correlation} &
0.47 &
0.48&
0.34&
0.59&
\textbf{0.69}\\

\hline
\end{tabular}
\caption{Assessment of the judges.}
\label{tab:judges}
\end{center}
\end{table}

We can notice that while the individual judges are relatively effective in judging the responses, the median answer of the three judges is the best performing option with a good level of correlation with the human judgment. Based on these results, we will assess the responses obtained with our large-scale evaluation presented in the next section using the mean value.

\section{Results} \label{sec:results}

This section describes the results obtained for each research question and discusses the potential threats to the validity of the results.

\begin{table}[ht]
\resizebox{\columnwidth}{!}{%
\begin{tabular}{l|l|l|l|l|l|l|}
\cline{2-7}
 & \textbf{p1} & \textbf{p2} & \textbf{p3} & \textbf{p4} & \textbf{p[1-4]} & \textbf{\begin{tabular}[c]{@{}l@{}}Never\\ found\end{tabular}} \\ \hline
\multicolumn{1}{|l|}{\textit{Retrieval: accurate question}} & 34\% & 4\% & 1\% & 2\% & 41\% & 5/25 \\ \hline
\multicolumn{1}{|l|}{\textit{Retrieval: inaccurate question}} & 10\% & 3\% & 2\% & 4\% & 19\% & 7/25 \\ \hline
\end{tabular}%
}
\caption{Probability of retrieving the correct block for accurate and inaccurate questions.}
\label{tab:questions_ranking}
\end{table}

\subsection{RQ1 - Accuracy}

This RQ investigates the effectiveness of both the intermediate retrieval task and the actual RAG's response. 

\subsubsection{RQ1.1 - Retrieval Accuracy}
To assess the accuracy of the retrieval process we measure how often the block that describes the troubleshooting procedure that must be actuated to respond to the question occurs in any of the first four blocks that are retrieved. In particular, Table~\ref{tab:questions_ranking} row \emph{Retrieval (accurate questions)} reports the probability that the correct block appears in any of the first four blocks (columns \emph{p1, p2, p3, p4}), the cumulative probability of appearing in any of these blocks (column \emph{p[1-4]}), and the number of questions for which the retrieval systematically failed, never retrieving the correct block (column \emph{never found}). 

In most cases, the correct block, if retrieved, is in the first position (34\% probability). Still, the other three positions are also often relevant, for a cumulative probability of retrieving the correct block equals 41\%. 
Although not always retrieving the correct description, when the retrieval fails, it can still retrieve related troubleshooting procedures that might help the LLM, and then the operator, to formulate a feasible troubleshooting plan, as confirmed by the results in the next section.
Further, in 80\% of the questions, the right block is retrieved at least in some of the repetitions, while for 5 questions (20\%) the right block could never be retrieved.

\subsubsection{RQ1.2 - Response Accuracy}

\begin{figure}[htb!]
    \centering
    \includegraphics[width=\linewidth]{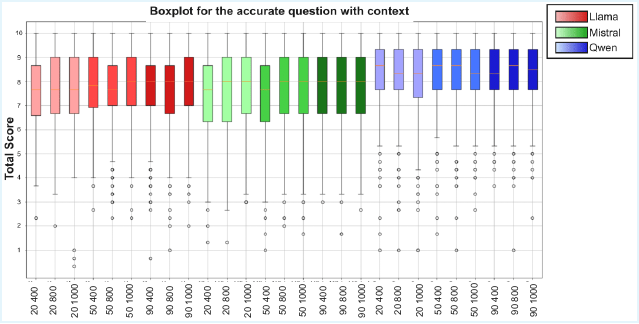}
    \caption{Boxplot representing the assessment of the answers to the accurate questions with context.}
    \label{fig:questCompleteCtx}
\end{figure}

Figure~\ref{fig:questCompleteCtx} reports the results obtained with the three LLMs when the accurate question with context is used. The color is representative of the model used, and the shade of the color differentiates configurations. 

We can notice how the models performed generally well according to the jury, with the boxplots all shown in the top part of the diagram. In particular, Qwen performed better than the others, both in terms of consistency of the results (shorter boxes) and in terms of the top scores reached. 

The configuration of the LLM (i.e., top-p and the chunk size) does not impact the results, suggesting that its choice is not particularly critical for the studied task.

\subsection{RQ2 - Sensitivity}

This RQ investigates the sensitivity to the accuracy of questions of both the intermediate retrieval task and the actual response. 

\subsubsection{RQ2.1 - Retrieval Sensitivity}
Table~\ref{tab:questions_ranking} row \emph{Retrieval (inaccurate questions)} reports the probability that the correct block appears in any of the first four blocks (columns \emph{p1, p2, p3, p4}), the cumulative probability of appearing in any of these blocks (column \emph{p[1-4]}), and the number of questions for which the retrieval systematically failed, never retrieving the correct block (column \emph{never found}), when an inaccurate question is used. 

The correct block occurs in the first position with only a 10\% chance. The cumulative probability of retrieving the correct block in any of the first four positions is 19\%, less than half of the probability observed when accurate questions are used. 
The use of synonyms and changes to the used verbs, due to the many similar procedures stored in the KB, immediately complicated the retrieval task.  

This result remarks how using the domain terminology as used in the troubleshooting procedure is important for the output of the retrieval task. Although the retrieval task is complicated by the inaccurate questions, the LLM may still have the chance of exploiting its knowledge and its generative capabilities to produce useful recommendations. 

\subsubsection{RQ2.2 - Response Sensitivity}

\begin{figure}[htb!]
    \centering
    \includegraphics[width=\linewidth]{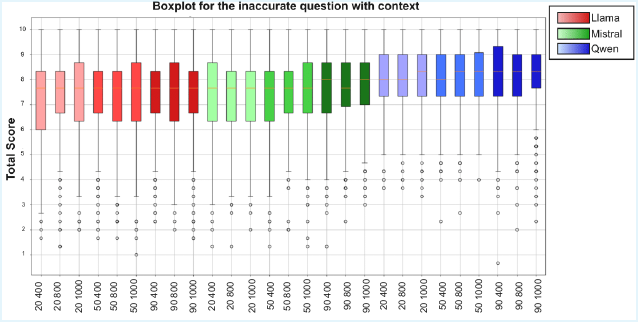}
    \caption{Boxplot representing the assessment of the answers to inaccurate questions with context.}
    \label{fig:shortCtx}
\end{figure}

\begin{figure}[htb!]
    \centering
    \includegraphics[width=\linewidth]{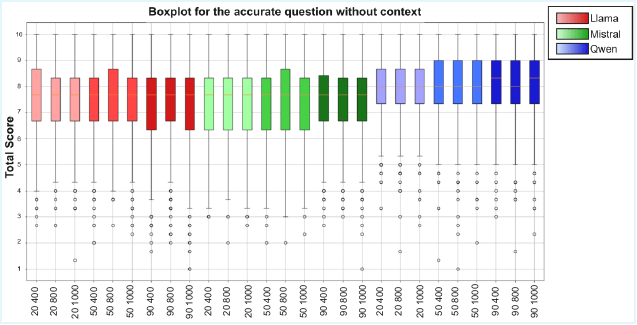}
    \caption{Boxplot representing the assessment of the answers to questions without context.}
    \label{fig:FullnoCtx}
\end{figure}

\begin{figure}[htb!]
    \centering
    \includegraphics[width=\linewidth]{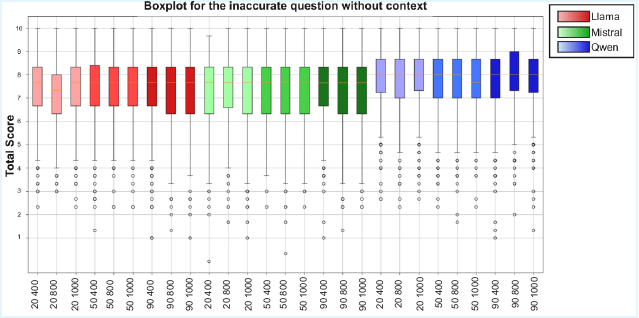}
    \caption{Boxplot representing the assessment of the answers to inaccurate questions with no context.}
    \label{fig:shortNoCtx}
\end{figure}

Figures~\ref{fig:shortCtx}, \ref{fig:FullnoCtx}, and \ref{fig:shortNoCtx} show the results for inaccurate questions with context, accurate questions without context, and inaccurate questions without context, respectively.

The box plots show two main results. First, results do not strongly degrade with the inaccuracy of the questions and the lack of context, according to the jury. Of course, both a nicely formulated question and a reference to the context seem useful to obtain a correct answer, but neither of the two looks indispensable. Second, the better performance of Qwen is consistently confirmed across all configurations. This is likely because Qwen is trained on a diverse multilingual corpus, and it demonstrates strength in instruction-following tasks, which makes it more robust to poorly formulated questions and context sparsity. 

It is interesting to note that Llama, although performing worse than Qwen, still performs well while using a smaller model. In fact, Llama is a 3B model, while Qwen is a 7B model. In case using a small model is important (e.g., due to resources on the available constraints), Llama might be a relevant option. 

Again, the choice of the configuration has little impact on the results.

\subsection{RQ3 - Derivation}
This RQ investigates whether the RAG can be used to derive troubleshooting procedures that are not already present in the KB, exploiting the generative capability of the LLM. To do this, we measure the loss of similarity between the answers obtained with the full KB and the answers obtained with the three alternative configurations, for all the considered LLMs. We use BLEU \cite{haque2022semantic} as a measure of similarity since it can capture the semantic closeness of two paragraphs.

\begin{table}[H]
\centering
\caption{Results with incomplete KB} \label{tab:incompleKB}
\begin{tabular}{lc c c }
\toprule
Scenario & no response & no entry & no KB\\

\midrule

$\Delta$\% BLEU (avg) &
\textcolor{red}{-9.68\%} &
\textcolor{red}{-53.30\%} &
\textcolor{red}{-85.83\%}\\

\bottomrule
\end{tabular}
\end{table}

Table~\ref{tab:incompleKB} shows the average loss across all questions. When only the response is missing from the KB, the LLM can still suggest the correct troubleshooting procedure in the vast majority of cases, with a loss smaller than 10\%. However, once the full entry with symptoms and the troubleshooting procedure is removed, the number of correct responses has a significant drop. Yet some questions could be answered by reusing and adapting troubleshooting procedures defined for other components and/or cases. This is a relevant capability that is useful to help operators address unspecified situations that require a timely resolution.

The drop is dramatic when the KB is empty, which, not surprisingly, indicates that the RAG is necessary to recommend troubleshooting procedures.

We further analyze these cases for the best configuration in the next section.

\subsection{RQ4-Qualitative Assessment}
To assess the results qualitatively, we inspected the outputs produced by Qwen, the best-performing model, in the context of RQ1 (accurate questions), RQ2 (alternative questions), and RQ3 (questions with no explicit answer in the KB), and analyzed the likely reasons for the mistakes. In particular, we categorized answers in three cases: \emph{all steps}, which are answers that include all the steps that must be executed to complete the troubleshooting procedures; \emph{partial}, which are answers that include only a subset of the steps that must be actuated; and \emph{wrong}, which are answers unrelated to the actual procedure. 
Table~\ref{tab:completeness} reports the results. 
For the all steps case, we also indicate in parentheses the percentage of cases where extra steps appear in the reported answer. 

When asking questions that have a response in the KB (row full KB), the qualitative analysis reveals that using proper questions and context is important to obtain a high rate of useful responses (only 8\% of the responses were wrong). The percentage of wrong responses grows to 36\% when either the question is inaccurate or the context is not provided. Finally, more than half of the responses are wrong for inaccurate questions without context. Results also show that the presence of the sources used to formulate the answer is important for validating the answer and resolving any imprecision, for instance, to retrieve the missing steps for the incomplete answers or filter out the extra steps.

When considering scenarios where operators ask questions about troubleshooting procedures that have not been documented yet, not surprisingly, the accuracy of the responses quickly drops. However,  it is an interesting result that for many cases (nearly one case out of two), the RAG was still able to output a useful response, sometimes a fully accurate procedure. In fact, if either the question is accurate or the context is provided, the range of wrong responses is from 52\% to 68\%.  The extreme scenario where the question is inaccurate, the context is not provided, and the response is not in the KB is definitely too hard for the RAG, with 72\% of the responses being wrong and 20\% incomplete. Although the RAG cannot be systematically used to derive troubleshooting procedures that are not in the KB, it can indeed be a valid tool to provide initial suggestions to the operator who has to react quickly to any problem.

We also inspected the responses to identify the elements that likely influenced the results.

When inspecting the answers generated for questions in RQ1, we noticed that accuracy is lower for the questions that show terminological ambiguity or include complex multi-condition formulations. Questions such as “\textit{What should I do if I detect compressed air leaks on pipelines?}” or “\textit{What should I do if the assembly of a component appears loose or uneven?}” use vague or generic terms like “pipelines” and “component” that may confuse the RAG. As a result, the system may fail to retrieve the most relevant chunks of information or retrieve unrelated content that does not fully address the issue.
Similarly, complex questions that include multiple conditions can introduce difficulties in both retrieval and generation. For instance, questions such as “\emph{What should I do if the compressor has been turned off and the error message 'Oil' is displayed on the compressor control?}” envisage the understanding of the interplay between two events, namely, the shutdown of the compressor and the appearance of a specific error message, not handling the two events has two independent failure triggering conditions. 
When the relevant information is scattered or only implicitly connected, the system may struggle to synthesize a coherent and correct response.
These issues highlight the importance of well-structured documentation and clear, unambiguous question phrasing when building and evaluating RAG systems.

When inspecting the answers generated for questions in RQ2, similarly to observations for questions in RQ1, a major problem lies in the use of ambiguous terminology. For example, in the question "\textit{What should I do if the CIP pump leaks?}", the verb "leaks" is too generic and does not clearly indicate whether the issue concerns the mechanical seals, pipe fittings, or some other components.
Here, the use of a vague term has likely negatively affected the capability of the model. 
Furthermore, the RAG has more difficulty in responding to questions that combine multiple conditions. A question such as "\emph{What should I do if, while the E/P is running, the pump caps are leaking?}" blends the operational status of the equipment with a symptom of failure. In such cases, the RAG system may not always find a suitable chunk that responds to the described scenario. Unclear phrasing, generic conditions, or overly complex formulations limit the effectiveness of the semantic retrieval, leading to less accurate answers.

When inspecting the answers generated for questions in RQ3, we noticed the LLM is often able to adapt troubleshooting procedures that are strongly similar to the missing one, often compensating for the lack of information in the KB.
Errors are mostly due to erroneous procedure reuses, erroneous combinations of multiple procedures into a single one, or the output of generic recommendations that do not consist of any procedure that operators can apply.

\begin{table*}[]
\resizebox{\textwidth}{!}{%
\begin{tabular}{l|lll|lll|lll|lll|}
\cline{2-13}
 & \multicolumn{3}{c|}{\textbf{Accurate with context}} & \multicolumn{3}{c|}{\textbf{Inaccurate with context}} & \multicolumn{3}{c|}{\textbf{Accurate without context}} & \multicolumn{3}{c|}{\textbf{Inaccurate without context}} \\ \cline{2-13} 
 & \multicolumn{1}{l|}{\textbf{All steps}} & \multicolumn{1}{l|}{\textbf{Partial}} & \textbf{Wrong} & \multicolumn{1}{l|}{\textbf{All steps}} & \multicolumn{1}{l|}{\textbf{Partial}} & \textbf{Wrong} & \multicolumn{1}{l|}{\textbf{All steps}} & \multicolumn{1}{l|}{\textbf{Partial}} & \textbf{Wrong} & \multicolumn{1}{l|}{\textbf{All steps}} & \multicolumn{1}{l|}{\textbf{Partial}} & \textbf{Wrong} \\ \hline
\multicolumn{1}{|l|}{\textit{full KB}} & \multicolumn{1}{l|}{64\% (28\%)} & \multicolumn{1}{l|}{28\%} & 8\% & \multicolumn{1}{l|}{48\% (12\%)} & \multicolumn{1}{l|}{16\%} & 36\% & \multicolumn{1}{l|}{44\% (20\%)} & \multicolumn{1}{l|}{20\%} & 36\% & \multicolumn{1}{l|}{24\% (12\%)} & \multicolumn{1}{l|}{20\%} & 56\% \\ \hline \hline
\multicolumn{1}{|l|}{\textit{no response}} & \multicolumn{1}{l|}{36\% (20\%)} & \multicolumn{1}{l|}{12\%} & 52\% & \multicolumn{1}{l|}{28\% (28\%)} & \multicolumn{1}{l|}{20\%} & 52\% & \multicolumn{1}{l|}{16\% (12\%)} & \multicolumn{1}{l|}{16\%} & 68\% & \multicolumn{1}{l|}{12\% (12\%)} & \multicolumn{1}{l|}{16\%} & 72\% \\ \hline
\multicolumn{1}{|l|}{\textit{no entry}} & \multicolumn{1}{l|}{28\% (20\%)} & \multicolumn{1}{l|}{12\%} & 60\% & \multicolumn{1}{l|}{40\% (4\%)} & \multicolumn{1}{l|}{20\%} & 64\% & \multicolumn{1}{l|}{36\% (8\%)} & \multicolumn{1}{l|}{12\%} & 56\% & \multicolumn{1}{l|}{20\% (0\%)} & \multicolumn{1}{l|}{20\%} & 72\% \\ \hline
\multicolumn{1}{|l|}{\textit{no KB}} & \multicolumn{1}{l|}{28\% (4\%)} & \multicolumn{1}{l|}{12\%} & 60\% & \multicolumn{1}{l|}{16\% (4\%)} & \multicolumn{1}{l|}{20\%} & 64\% & \multicolumn{1}{l|}{32\% (0\%)} & \multicolumn{1}{l|}{12\%} & 56\% & \multicolumn{1}{l|}{8\% (4\%)} & \multicolumn{1}{l|}{20\%} & 72\% \\ \hline
\end{tabular}%
}
\caption{Results of the manual inspection of the best-configuration's results in terms of completeness.}
\label{tab:completeness}
\end{table*}

\subsection{RQ5 - Performance} 
Since operators are supposed to query the RAG during problematic situations, we measured the time required by the RAG to provide an answer to the questions formulated by users.

In Table~\ref{tab:llm_time}, we reported the time in seconds required by the RAG to provide an answer to questions w.r.t. the considered LLMs. Specifically, we reported for every LLM (i) the average (AVG) time in seconds taken by the RAG to answer the questions considered in the advanced test; (ii) the fastest time (MIN) and (iii) the slowest (MAX) one. Looking at the reported times, Llama and Mistral can respond faster than Qwen, producing an answer in the range from about 2 to 7 seconds. 
Instead, Qwen, which is the most effective model in terms of accuracy and sensitivity, is slower in producing answers, responding in an average time of about 12 seconds. 

\begin{table}[ht]
    \centering
    \begin{tabular}{c|c|c|c}
    \hline 
      \textbf{LLM} & \textbf{AVG} & \textbf{MIN} & \textbf{MAX} \\  \hline
      Llama &	3,6 secs &	2,9 secs	& 4,2 secs \\
    Mistral &	4,52 secs &	3,5 secs	& 7,5 secs \\
    Qwen &	12,2 secs &	10,5 secs	& 14,2 secs \\ \hline
    
    \end{tabular}
    \caption{Time required by the RAG to provide an answer to questions w.r.t. the considered LLMs.}
    \label{tab:llm_time}
\end{table}

\smallskip

In a nutshell, in case the priority is obtaining responses quickly (e.g., in a few seconds), because a procedure must be immediately selected, or the answer is the basis for some follow-up steps (e.g., additional questions), Llama might be the preferred option. Instead, if obtaining accurate answers is the priority, at the cost of waiting some more time (between 9 and 10 secs in our experiments), Qwen is the best choice among the investigated LLMs.

\subsection{Threats to Validity}
In this section we discuss potential threats to the validity of the conducted evaluation.

\textbf{Construct Validity.} Using LLM-based judges to assess the accuracy of generated answers may generate a concern related to their judgment, which might reflect limitations or inconsistencies in their capabilities. To mitigate this threat, we completed an initial assessment of the judges in our context (see Section~\ref{sec:judges}), identifying the combination that provides the best result. Further, we manually inspected the responses, producing an accurate qualitative assessment of the best-performing model, eliminating the issue of judges for the most promising configuration of the approach.
\review{R3}{Another aspect that may affect reproducibility is version drift of the used LLMs. Our evaluation based on multiple LLMs reduces the dependency of the results on a single model version.}

\textbf{Internal Validity.} In our setup, we study the impact of three parameters (LLM, chunk size, top-p) on the results, partially considering their interactions. Further experiments would be necessary to systematically consider every possible combination of these parameters, although the evidence reported in this paper suggests that their impact on the results is very limited.

While we designed question variants to represent different levels of precision and contextual reference, there might be unintended differences in difficulty or phrasing that influence the outcome beyond what we intended to measure. To mitigate this threat, we discussed cases of questions that have been hard to answer, so that these could be taken into consideration in future research.

\textbf{External Validity.} Our experiments are based on a set of 25 distinct topics and 100 questions, which, despite being diverse and relevant in the studied domain, cannot represent the full variety of real-world user queries. The study in an industrially relevant context posed limitations in the range and number of cases that could be investigated. Yet our findings  could be the basis for follow-up work in similar domains.

\textbf{Conclusion Validity.} The assignment of a numerical score to an answer provides granularity, but it might not capture all dimensions of answer quality. 
For this reason, the paper reports a detailed qualitative analysis of the responses. 
\review{R2}{Although, we did not involve practitioners from Fincantieri in this phase due to organizational and operational constraints, we cross-validated and reviewed the interpretations to reduce bias and ensure consistency. }

\section{Lesson Learned} \label{sec:lesson}
Finally, we discuss the lessons learned from our experiments.

\textbf{Lesson 1 – RAG is a valid tool to suggest troubleshooting procedures, which, however, must be validated and completed by experts.} RAG revealed a useful tool to provide useful recommendations that contain all the steps, or at least some of the steps, that must be actuated to resolve an incident (see row Full KB in Table~\ref{tab:completeness} column \emph{accurate with context}). As in many other contexts, operators cannot blindly trust generative AI, and the operator's intervention is required to critically revised the procedure and finally decide the steps to execute. In this respect, the set of sources attached to the responses is a valid support to quickly cross-validate the responses.

\textbf{Lesson 2 – Training operators on the domain terminology might improve the capability to quickly respond to incidents.} Results show that some responses might be imprecise, especially when queries are inaccurate and lack contextual information (see row Full KB in Table~\ref{tab:completeness} column from second to fourth). A consequent recommendation is to train operators on the main terms that must be used when looking for troubleshooting procedures, so that they can formulate proper requests for the RAG.

\textbf{Lesson 3 - Accuracy is resilient to question-inherent variations but sensitive to ambiguity.} 
The experiments demonstrate that LLMs like Qwen often maintain good performance even when questions are poorly formulated or lack explicit context. However, terminological ambiguities and complex multi-condition questions still pose challenges, leading to decreased response accuracy. Ensuring question clarity and minimizing vagueness can significantly improve system reliability. This issue can likely be mitigated by employing prompt engineering techniques or implementing strategies to automatically rewrite prompts.

\textbf{Lesson 4 – The generative capability of RAGs may help face undocumented situations.}  
The generative capabilities of RAGs, and LLMs in particular, allow them to infer some plausible troubleshooting steps even when procedures are missing or incomplete in the KB (see rows from second to fourth in Table~\ref{tab:completeness}). Although this behavior is promising for handling undocumented scenarios, it also introduces the risks of hallucination or overgeneralization, especially when questions are not properly formulated. In safety-critical environments, such responses must be carefully validated, and systems should be designed to clearly distinguish between reliably retrieved content and unreliable inferred content.

\textbf{Lesson 5 – Smaller models offer speed, larger models offer accuracy.}  
Our experiments reveal a trade-off between response time and response quality (Table~\ref{tab:llm_time}). Smaller models like Llama respond faster and may be suitable for time-sensitive tasks, while larger models like Qwen provide more accurate and context-aware responses at the cost of increased latency. This suggests that the choice of model should be guided by the operating context, depending on whether speed or accuracy is the priority.

\textbf{Lesson 6 - Semantic search in isolation is weaker than RAG.} The semantic retrieval task was able to retrieve the correct block a relatively high number of times (Table~\ref{tab:questions_ranking}). In our experiments, the role of the LLM was essential to derive the troubleshooting procedure, compensating for some inaccuracies in the retrieval task. In addition, the RAG offers a conversational interface that is particularly effective for operators handling critical situations. 
\section{Related Work} \label{sec:related}

Multiple classes of approaches have been defined to address failure and failure symptoms. For instance, systems might be designed to be resilient to failures \cite{Mingyue:ByzantineFailure:2022,Kevin:ResilientAutonomousVehicles:2022,mariani2020predicting,Campos:FailurePrediction:2023}, so that failures have no harmful or critical consequences \review{R1}{especially in sensitive domains such as healthcare \cite{Amugongo2025}}. 
Other approaches, not designed for resiliency, studied how to exploit the implicit redundancy present in some systems to dynamically generate workarounds~\cite{10.1145/2755970}. Finally, once a failure has been observed, automatic program repair techniques have been investigated to recommend fixes to developers~\cite{8089448,10.1145/3631974}.
Complemental to these studies, this paper investigates the case systems cannot workaround failures automatically, but require the user intervention to react to failure symptoms.

Our work adapts the principles of Recommendation Systems for Software Engineering (RSSE) \cite{robillard2009recommendation,gasparic2016recommendation} to a largely unexplored domain in the operational maintenance of CPSs. RSSEs are designed to overcome information overload and support users in decision-making and information-seeking activities by providing valuable information items for a specific software engineering task within a given context \cite{robillard2009recommendation}. While recommendation systems in software engineering typically assist software developers by suggesting source code artifacts during the implementation and maintenance phases of the development lifecycle \cite{di2021development,gasparic2016recommendation}, our system supports operators during the post-deployment phase. Instead of recommending code, it provides troubleshooting procedures extracted from technical manuals to help resolve runtime failures. This shift in focus from code to documentation addresses a recognized gap in the literature, where there are unexploited opportunities in the development of recommendation systems outside the source code domain \cite{robillard2009recommendation,gasparic2016recommendation}. By leveraging RAG, our approach broadens the applicability of recommendation systems to support decision-making in the resolution of runtime failures within industrial CPS.

Recent advancements in 
troubleshooting within industrial environments have been driven by the integration of natural language processing and information retrieval techniques. These studies have explored these avenues to enhance accessibility, efficiency, and accuracy of maintenance operations~\cite{Alfeo:troubleshooting:JIM:2021,Su:Troubleshooting:Access:2019,Ren:RobotMaintenance:Access:2024}.

Ren et al. \cite{Ren:RobotMaintenance:Access:2024} developed a voice-interactive fault diagnosis system for industrial robots, which demonstrates the potential of voice commands for facilitating information retrieval from extensive manuals. While they focus on voice interaction, our study enhances this by applying RAG techniques to improve the accuracy and responsiveness of troubleshooting information retrieval.

Kiangala and  Wang \cite{kiangala2024experimental} present an experimental hybrid AI chatbot that combines customized AI and generative AI to enhance human–machine interaction within factory troubleshooting scenarios under Industry 5.0. Their approach emphasizes adaptive, AI-driven dialogue systems capable of resolving complex troubleshooting queries, leading to reduced factory downtime. Our work contributes to this field by demonstrating how RAG can enhance the retrieval of relevant data, thus improving the interaction between humans and machines in industrial settings.

Su et al. \cite{Su:Troubleshooting:Access:2019} introduce an innovative approach to enhance maintenance and troubleshooting efficiency in aerospace applications. Their method leverages integrated information flow modeling and ontology, along with ant colony optimization, to detect faults. 
Unlike Y. Su et al.'s method, which relies on structured data and ontological frameworks, our approach leverages LLMs to interpret unstructured queries. This enables operators to intuitively engage with procedures without needing to construct ad hoc ontologies.

The findings presented in the work by Algeo et al.~\cite{Alfeo:troubleshooting:JIM:2021} emphasize the critical need for integrated maintenance strategies in industrial settings. The study uses a deep learning approach for retrieving solutions from historical technical assistance reports within predictive maintenance frameworks, 
significantly reducing downtime and improving operational efficiency. The integration of such technologies aligns with our research focus, as it supports the need for robust data-driven decision-making processes that can be complemented by RAG techniques to improve information retrieval.

\section{Conclusions} \label{sec:conclusions}
Large industrial CPSs are complex systems that require operators to readily resolve unexpected behaviors and failures. The troubleshooting strategies that must be applied are often documented in natural language manuals that are hard and slow to search. 

In this paper, we investigated the use of RAG to quickly identify the right procedure to resolve a problem. We studied this problem for the manuals documenting a large naval CPS available at Fincantieri. Results show that RAG can indeed help operators, with some caveats, such as the documentation of the sources used to produce answers. From our experience, we distilled a lesson learned that can be used as a foundation for additional work in the area. 

In the future, we plan to extend our work to additional systems and to develop additional strategies to help operators using conversational interfaces to timely react to problems. \review{R3}{This way, we want to test other CPS documentation in order to explore how RAG responds to multi-modal troubleshooting instructions.}
\begin{acks}
This work was supported by the ATOS project, funded by the MUR under the PNRR- CN -
HPC - ICSC program (CUP: H43C22000520001).
\end{acks}

\bibliographystyle{ACM-Reference-Format}
\bibliography{main}


\end{document}